\documentclass[prl,preprint,superscriptaddress]{revtex4}
\usepackage{graphicx} 
\usepackage{bm} 
\usepackage{multirow} 
\usepackage{subfigure}
\usepackage{amsmath,amsfonts}
\usepackage{color}

\begin{document}
 
\title{Effects of Cutoff Functions of Tersoff Potentials on Molecular Dynamics Simulations of Thermal Transport}
 
\author{X. W. Zhou}
\email[]{X. W. Zhou: xzhou@sandia.gov}
\affiliation{Mechanics of Materials Department, Sandia National Laboratories, Livermore, California 94550, USA}

\author{R. E. Jones}
\affiliation{Mechanics of Materials Department, Sandia National Laboratories, Livermore, California 94550, USA}

\date{\today}
 
\begin{abstract}

Past molecular dynamics studies of thermal transport have predominantly used Stillinger-Weber potentials. As materials continuously shrink, their properties increasingly depend on defect and surface effects. Unfortunately, Stillinger-Weber potentials are best used for diamond-cubic-like bulk crystals. They cannot represent the energies of many metastable phases, nor can they accurately predict the energetics of defective and surface regions. To study nanostructured materials, where these regions can dominate thermal transport, the accuracy of Tersoff potentials in representing these structures is more desirable. Based upon an analysis of thermal transport in a GaN system, we demonstrate that the cutoff function of the existing Tersoff potentials may lead to problems in determining the thermal conductivity. To remedy this issue, improved cutoff schemes are proposed and evaluated.

\end{abstract}
 


\maketitle


Thermal conductivity of semiconductors is often the limiting factor for important technological applications. For instance, a higher thermal conductivity improves heat dissipation, which also enables an increase in the microelectronic device density \cite{S2006}. On the other-hand, a lower thermal conductivity has long been sought in order to push the energy conversion efficiency of thermoelectric devices beyond a critical high value \cite{MSS1997}. As the system dimensions continuously shrink towards the nanoscale, the thermal conductivity increasingly depend on defects and surfaces of materials. Because experimental study is challenging at the nanoscale, a theoretical understanding of the influence of defects and surfaces on thermal conductivity is essential for optimizing nanoscale applications. With the structures and energetics of defects and surfaces precisely captured in a computational lattice, molecular dynamics (MD) has emerged as a powerful means to study the thermal transport in nano crystals. MD has been successfully applied for both covalent (semiconductors) \cite{zKKK2005,VC2000,MMP1997,OS1999,SPK2002,SPK2004,WLXO2009,ZAJGS2009,ZJA2009,ZJA2010,PSM2007,CCDG2000,CCG2000,LPY1998} and other (e.g., ionic) \cite{LMH1986,VHC1987,M1992,PB1994,B1996,JJ1999,YCSS2004,SP2001} systems.

MD simulations of thermal transport in semiconductors have used predominantly the Stillinger-Weber \cite{SW1985} rather than the Tersoff/Brenner \cite{T1988,T1989,B1990} types of potentials. Of the papers impartially cited here, a majority \cite{zKKK2005,VC2000,MMP1997,OS1999,SPK2002,SPK2004,WLXO2009,ZAJGS2009,ZJA2009,ZJA2010,PSM2007} applied Stillinger-Weber potentials and only a few \cite{CCDG2000,CCG2000,LPY1998} used Tersoff/Brenner types of potentials. Recently, we published a series of papers on MD simulations of thermal transport in GaN \cite{ZAJGS2009,ZJA2009,ZJA2010}. While these studies employed a Stillinger-Weber GaN potential \cite{BS2002,BS2006}, significant efforts were also made to explore the Tersoff GaN potential \cite{NAEN2003} during the course of the work. With this particular parameterization of the Tersoff potential, we found it difficult to ensure energy conservation during a ``constant energy'' MD simulation unless the time step size was significantly reduced from the 0.001 ps we normally used for other potentials, and the thermal conductivities obtained were more than one order of magnitude smaller than the values we would otherwise obtain from the Stillinger-Weber potential \cite{BS2002,BS2006}. 

From a physics point of view, Stillinger-Weber potentials are empirically designed for diamond-cubic-like bulk environment and they do not work well in defective and surface regions. For instance, the angular term in Stillinger-Weber potential is essentially a positive parabolic energy penalty for deviations of bond angles from the tetrahedral angle. As this positive energy term is repulsive, any structures with non-tetrahedral bond angles would have longer bond lengths than those in diamond-cubic-like crystals or dimers. This is not always true in reality. Furthermore, under the condition that the diamond-cubic-like crystal has the lowest energy, the angular term in Stillinger-Weber potential must be constrained above a certain level. As a result the potential cannot correctly describe the trend of energies in many other metastable phases. Tersoff potentials, on the other hand, inherit the quantum mechanics-based bond order concept \cite{PFNMZW2004,DMNZWP2005,MZWNDP2006}. They are therefore well suited for treating defects and surfaces. In particular, we note that the GaN Tersoff potential \cite{NAEN2003} was carefully parameterized using cohesive energies, atomic volumes, defect properties, melting temperatures, and solubility limits of a variety of stable and metastable phases. Its fidelity has also been validated in MD simulations of vapor deposition that correctly showed the crystalline growth of the lowest energy phase \cite{ZMGW2006}.

The energy non-conservation and the abnormally low thermal conductivity are not transparent problems. Intuitively, they cannot be attributed directly to the potential format. Resolving these significant problems is important to the materials modeling community, as it will both promote the use of Tersoff potential in future MD simulations and improve the atomistic simulations of semiconductor systems. Understanding the abnormally low thermal conductivity predicted by the GaN Tersoff \cite{NAEN2003} potential may also result in the discovery of a new thermal scattering phenomenon that can be utilized to engineer thermal conductivity of materials. This paper will address these issues.

Following the format of the GaN paper \cite{NAEN2003}, the Tersoff potential is expressed as
\begin{equation}
E = \sum_{i}\sum_{j>i}\left[\phi_{ij}\left(r_{ij}\right)-\bar{B}_{ij}\cdot\psi_{ij}\left(r_{ij}\right)\right]
\label{energy}
\end{equation}
where the bond order $\bar{B}_{ij}$ is defined as 
\begin{equation}
\bar{B}_{ij} = \frac{\sqrt{1+\chi_{ij}}+\sqrt{1+\chi_{ji}}}{2}
\label{bond order}
\end{equation}
with the local angular-dependent variable $\chi_{ij}$ calculated as
\begin{equation}
\chi_{ij} = \sum_{k\neq i,j}f_{ik}\left(r_{ik}\right)\cdot g_{ik}\left(\theta_{jik}\right)\cdot \exp\left[2\mu_{ik}\left(r_{ij}-r_{ik}\right)\right]
\label{chi}
\end{equation}
Eqs. (\ref{energy}) and (\ref{chi}) involve three pair functions $\phi_{ij}$, $\psi_{ij}$, and $f_{ik}$ representing respectively the i-j repulsion, the i-j attraction, and the i-k cutoff (in the angularly dependent $\chi$ term). Here the symbol $r_{ij}$ represents the usual relative distance between atoms $i$ and $j$ and $\theta_{jik}$ is the angle between the triplet $jik$ with $i$ at its vertex. The cutoff function $f$ is taken to be  a sinoidal spline: 
\begin{eqnarray} 
f_{ij}\left(r_{ij}\right) = \left\{
\begin{array}{ll} 
1, & r_{ij} \leq R_{ij}-D_{ij} \\ 
\frac{1}{2} - \frac{1}{2} \sin\left[\frac{\pi\left(r_{ij}-R_{ij}\right)}{2D_{ij}}\right], & R_{ij}-D_{ij} < r_{ij} \leq R_{ij}+D_{ij} \\
0, & r_{ij} > R_{ij}+D_{ij} 
\end{array}\right.
\label{f} 
\end{eqnarray} 
Here $r_s = R-D$ is the cutoff starting radius and $r_c = R+D$ is the cutoff distance at which point the interactions go to zero. To impose the cutoff of the $\phi$ and $\psi$ functions, the Tersoff potentials simply define $\phi_{ij}(r_{r_ij}) =  f_{ij}(r_{ij})\cdot V_{ij}^R(r_{ij})$ and $\psi_{ij}(r_{r_ij}) =  f_{ij}(r_{ij})\cdot V_{ij}^A(r_{ij})$, where $V^R(r)$ and $V^A(r)$ are two exponentially decaying functions \cite{NAEN2003,T1988,T1989,B1990} that are not subject to the discussions here. Due to the use of Eq. (\ref{f}), the three functions $\phi$, $\psi$, and $f$ have continuous values and the first derivatives, but their second derivatives abruptly change at the junction points $r_s$ and $r_c$. This is contrary to the Stillinger-Weber potentials whose high order derivatives are always continuous. 

\begin{figure}
\includegraphics[width=3in]{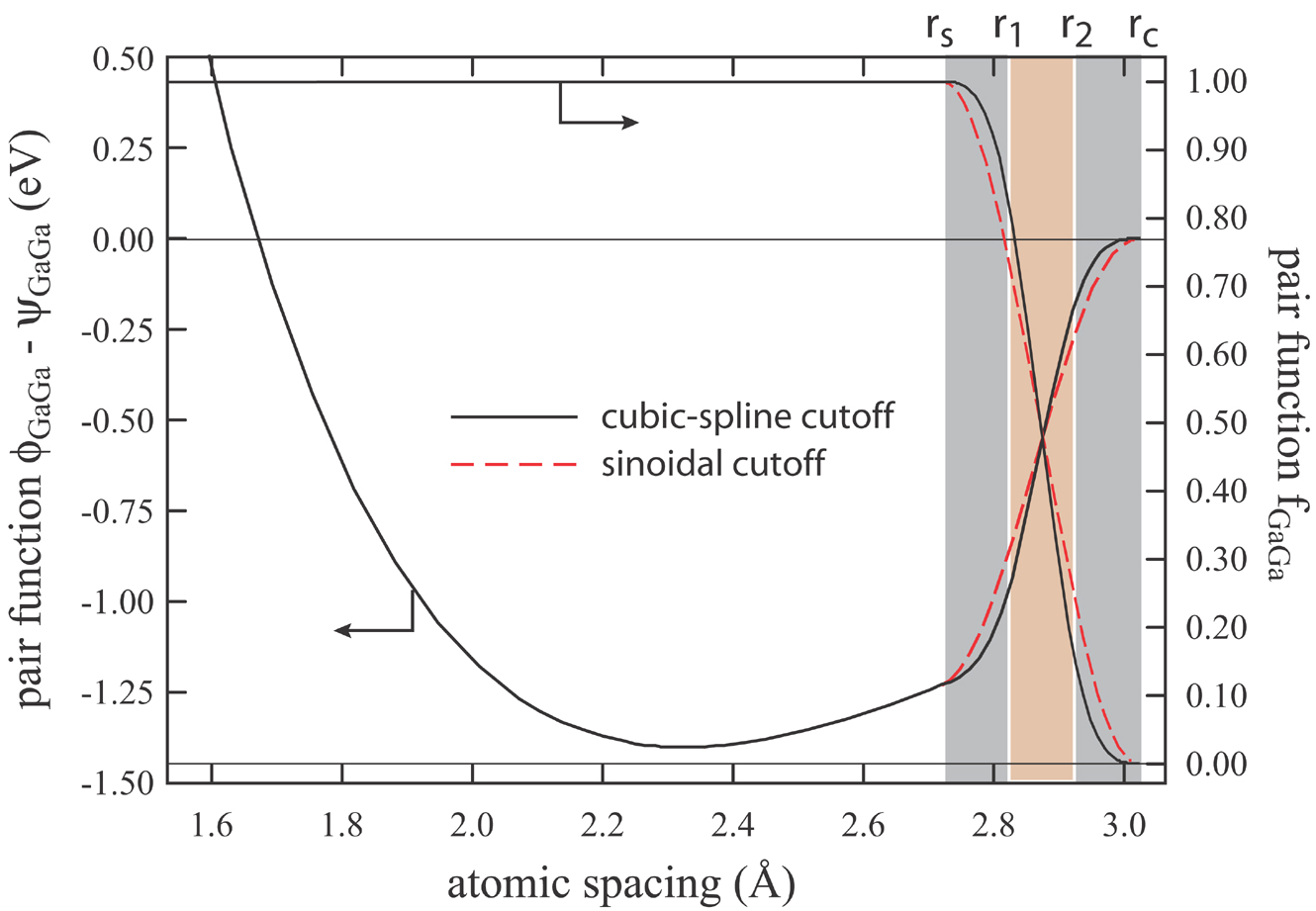}
\caption{GaGa pair interaction functions ($\phi_{GaGa}$ - $\psi_{GaGa}$ and $f_{GaGa}$). 
\label{GaGa cutoff function}}
\end{figure}

By dividing the $r_s$ - $r_c$ range evenly into three sections with two internal points $r_1, r_2 \in (r_s,r_c)$, a cubic spline cutoff approach can be used to remove the discontinuous 2nd-derivatives of the $\phi$, $\psi$, and $f$ functions. Using $\phi$ as an example, we can write,
\begin{eqnarray} 
\phi_{ij}\left(r_{ij}\right) = \left\{
\begin{array}{ll} 
V^R_{ij}\left(r_{ij}\right), & r_{ij} \leq r_{s,ij} \\ 
\sum_{n=0}^{n=2} \frac{{V^R_{ij}}^{(n)}\left(r_{s,ij}\right)}{n!}\left(r_{ij}-r_{s,ij}\right)^n + a_3 \left(r_{ij}-r_{s,ij}\right)^3, & r_{s,ij} < r_{ij} < r_{1,ij} \\
\sum_{n=0}^{n=3} b_n \left(r_{ij}-r_{2,ij}\right)^n, & r_{1,ij} \leq r_{ij} \leq r_{2,ij} \\
c_3 \left(r_{ij}-r_{c,ij}\right)^3, & r_{2,ij} < r_{ij} \leq r_{c,ij} \\
0, & r_{ij} > r_{c,ij}
\end{array}\right.
\label{cubic spline} 
\end{eqnarray} 
Eq. (\ref{cubic spline}) is continuous up to the second derivative at $r_s$ and $r_c$. It involves six parameters $a_3$, $b_0$ - $b_3$ and $c_3$ that can be solved by requiring the values as well as the first and the second derivatives to be continuous at the two added junction points $r_1$ and $r_2$. For an example comparison, the modified and unmodified $\phi_{GaGa} - \psi_{GaGa}$ and $f_{GaGa}$ functions are shown as the black and red dash lines respectively in Fig. \ref{GaGa cutoff function}. Note that equating the bond energy with $\phi_{GaGa} - \psi_{GaGa}$ implies a bond order of $\bar{B}_{GaGa} = 1$. This does not represent the real scenario but is sufficient for the discussion here. It can be seen that the modified potential does not significantly differ from the original potential. In fact, Eq. (\ref{cubic spline}) should be applicable to any existing Tersoff potentials without a need to re-parameterize. This cubic spline cutoff scheme has been implemented in the MD simulator LAMMPS \cite{P2008}.
 
\begin{figure}
\includegraphics[width=3in]{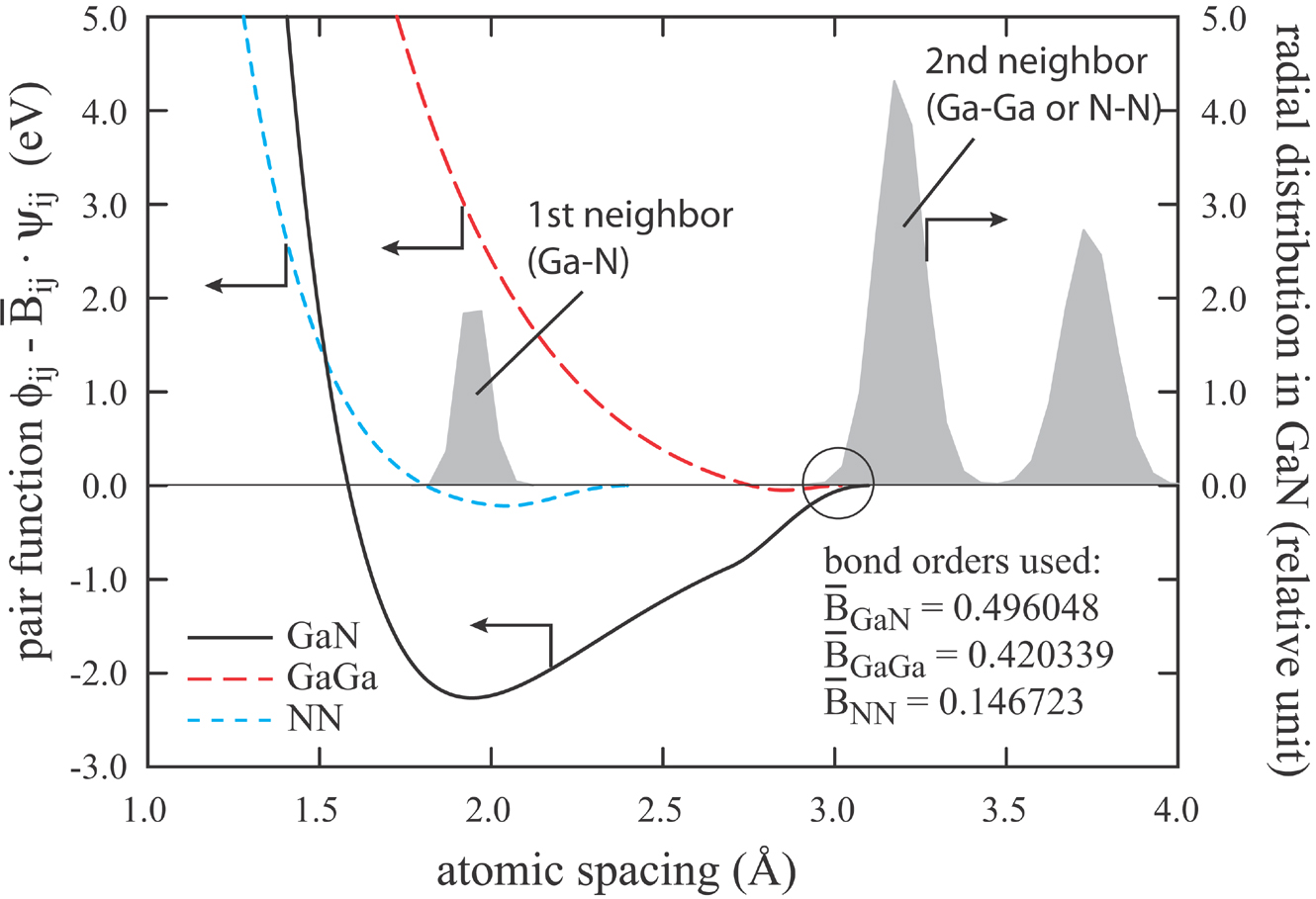}
\caption{Different pair interactions in wurtzite GaN crystal. 
\label{cutoff analysis}}
\end{figure}

We suspect that the discontinuous 2nd-derivatives are likely to be the source of the problems with the Tersoff potentials. To explore this, we calculate the bond energy $\phi_{ij}\left(r_{ij}\right)-\bar{B}_{ij}\cdot\psi_{ij}$ as a function of bond length $r_{ij}$ for different bonds ij = GaN, GaGa, NN using a fixed bond order $\bar{B}_{ij}$ estimated from the equilibrium wurtzite GaN and Eq. (\ref{bond order}). The results are shown in Fig. \ref{cutoff analysis} using black, long red dash, and short blue dash curves to represent, respectively, the Ga-N, Ga-Ga, and N-N interactions. The radial distribution of atoms was also determined from a GaN wurtzite equilibrated at a temperature of 300 K, and the result is included in Fig. \ref{cutoff analysis} as the shaded area. Fig. \ref{cutoff analysis} indicates that the third- or higher- nearest neighbors do not contribute to material properties as the corresponding radial distributions do not fall within ranges of any of the three interaction curves. Similarly, the N-N interaction does not contribute to properties because the N-N energy curve only overlaps with the nearest neighbor radial distribution which is between Ga and N. In fact, only the nearest neighbor Ga-N and the second-nearest neighbor Ga-Ga interactions contribute to properties. However, because the cutoff starting point $r_{s,GaN}$ (= 2.7 \AA~\cite{NAEN2003}) is well above the upper limit of the first radial distribution peak, how the Ga-N interaction is truncated has no effect on properties. Clearly, the effect due to a discontinuous 2nd-derivative, if any, must come from the Ga-Ga interaction.

In addition to the cubic spline cutoff modification, Fig. \ref{cutoff analysis} suggests numerous other ways to remove or mitigate the discontinuous 2nd-derivatives of the functions. As indicated by the circle in the figure, the Ga-Ga interaction has an insignificant contribution to the energy near the cutoff distance. In fact, there is no energy contribution at all at 0 K because the second neighbor (Ga-Ga) distribution would then reduce to a single line at approximately the peak position outside of the Ga-Ga interaction range. This means that neglecting the Ga-Ga interactions would only cause a negligible energy change and would not alter the relative phase stabilities. Similarly, a reduction of the cutoff starting point $r_{s,GaGa}$ (at a constant $r_{c,GaGa}$ cutoff distance) also does not significantly change the energy. A reduced $r_{s,GaGa}$, however, increases the cutoff transition range and therefore can be used to reduce magnitude of the 2nd-derivative near the transition. When no Ga-Ga interactions are considered, the only remaining interactions are between Ga and N, whose radial distribution is near the Ga-N interaction energy minimum which is far away from the cutoff distance. Consequently, the Ga-N neighbors do not change during thermal transport simulations where no phase transformation occurs. Hence, we can initialize the Ga-N neighbors at the start of simulations and not perform reneighboring calculations during the rest of simulations. It should be noted that the purpose of cutoff in atomistic simulations is to define the neighbor lists. When the correct Ga-N neighbors are initialized and are not recalculated, there is really no need to apply the cutoff algorithm to the Ga-N interaction functions. Hence, we can simply set $f_{GaN} \equiv 1$ to remove the discontinuous 2nd-derivatives of the Ga-N interactions. 

Based upon these observations, several schemes (as illustrated in Table \ref{results}) are designed to test the effects of the discontinuous 2nd-derivatives of the Ga-Ga functions near their cutoffs: A, the original potential without modifications (time step size 0.0010 ps unless indicated otherwise); B, the original potential simulated at a reduced time step size of 0.0005 ps; C, a reduced $r_{s,GaGa}$ value to reduce the 2nd-derivatives (and hence the extent of the discontinuity) of the Ga-Ga pair functions at $r_{c,GaGa}$; D, with only Ga-N interactions that are initialized without reneighboring; E, same as D with a further condition of $f_{GaN} \equiv 1$, and F, the cubic spline modification of the potential with other conditions the same as in scheme A. All these schemes were explored in thermal transport simulations.
\begin{table}
\caption{Results of temperature drift $\Delta T$ (K) and thermal conductivity $\kappa$ ($W/K \cdot m$) obtained from MD simulations over a 10 ns period using different combinations of GaGa cutoff function, GaGa cutoff starting point $r_{s,GaGa}$ (\AA), reneighboring, and time step sizes dt (ps).}
\begin{ruledtabular}
\begin{tabular}{ccccccc}
schemes&A&B&C&D&E&F\\\hline  
cutoff function&sinoidal&sinoidal&sinoidal&sinoidal&none (f $\equiv$ 1)&cubic spline\\\hline 
$r_{s,GaGa}$&2.72&2.72&2.32&2.72&N/A&2.72\\\hline 
reneighboring&yes&yes&yes&no&no&yes\\\hline
dt&0.0010&0.0005&0.0010&0.0010&0.0010&0.0010\\\hline
$\Delta$T&27&3&23&3&$<$1&$<$1\\\hline
$\kappa$&5.4$\pm$0.1&6.2$\pm$0.04&19.0$\pm$0.3&13.9$\pm$0.1&14.6$\pm$0.1&12.1$\pm$0.1\\
\end{tabular}
\end{ruledtabular}
\label{results}
\end{table}

\begin{figure}
\includegraphics[width=3in]{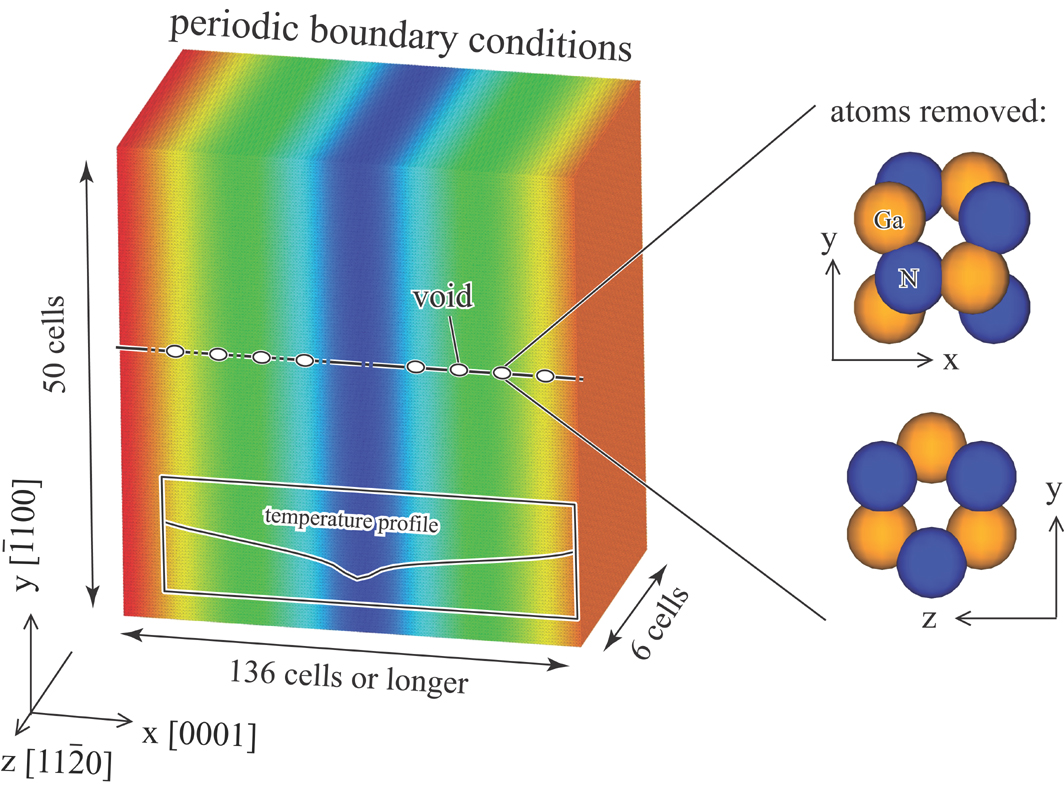}
\caption{Atomistic configuration for MD simulations.
\label{MD model}}
\end{figure}

The ``direct method'' MD approach \cite{ZAJGS2009,ZJA2009,ZJA2010} was employed to calculate the thermal conductivities of GaN wurtzite crystal along the $[0001]$ direction. The computational system used for the simulations is shown in Fig. \ref{MD model}, where the red and blue colors indicate respectively the hot and cold regions. The systems are 136, 50, and 6 cells in the x-, y-, and z- directions, corresponding approximately to a 708$\times$276$\times$18 \AA$^3$ material volume. Four voids are evenly created between the hot and the cold regions by removing six Ga and six N atoms as shown in Fig. \ref{MD model}. All simulations were carried out at a temperature of 300 K using periodic boundary conditions in all three coordinate directions. The simulated system therefore represents an infinite GaN crystal containing uniform defects. Thermal conductivities were averaged over 10 ns. The resulting temperature drift $\Delta T$ (energy drift) and thermal conductivity are compared as a function of simulation conditions in Table \ref{results} for various schemes. 

Table \ref{results} indicates that all simulations that involve discontinuous 2nd-derivatives of the Ga-Ga interactions, such as schemes A and C, have an energy conservation problem at a time step size of 0.0010 ps (the temperature drift during the 10 ns simulation is $\Delta T$ $>$ 20 K). This finding agrees well with the previous result \cite{ESD2005} that non-smoothness of the potential functions causes energy drift. Reducing the time step to 0.0005 ps can significantly improve the energy conservation, as in scheme B. The thermal conductivities obtained from simulations with large energy drift may not be accurate. Nonetheless, Table \ref{results} clearly shows that the finite 2nd-derivatives of the Ga-Ga functions at the cutoff distance also contribute to the abnormally low thermal conductivities, as with schemes A and B. Reducing the 2nd-derivatives by reducing the cutoff starting point $r_{s,GaGa}$ increases thermal conductivity, as in scheme C. All simulations that remove the discontinuous 2nd-derivatives, schemes D - E, have good energy conservation and reasonably large thermal conductivities. Note that both schemes D and E ignore the Ga-Ga interactions and use the Ga-N neighbors determined at the start of simulations. Scheme E further assumes $f_{GaN}(r_{GaN}) \equiv 1$. However, this change of the Ga-N cutoff does not have a significant effect because the Ga-N spacing of the crystal is well below the cutoff starting point $r_{s,GaN}$ as mentioned above. As a result, schemes D and E are expected to produce similar results. Because the reduction in thermal conductivity due to non-smooth cutoff function is an unphysical artifact, and because scheme E does not apply any cutoff modifications, the good agreement between the results of schemes E and F strongly validates the cubic spline cutoff approach.

In summary, we discovered that the existing Tersoff potential formats do not always produce the correct results due to the discontinuous 2nd-derivatives of the cutoff functions. This problem can be solved by applying the cubic spline cutoff modifications. In some thermal transport simulations where atomic neighbors do not change, one can initialize the neighbor lists at the start of simulations and not perform reneighboring. This is not only computationally efficient, but also eliminates the need to apply the cutoff algorithms to the functions and thereby removes the problem of the Tersoff potential. When the discontinuous 2nd-derivatives come from an insignificant second neighbor interaction, the problem can be simply solved by ignoring that interaction. Finally, we discovered the fundamental fact that the finite 2nd-derivatives of the interaction between atoms near the cutoff distance can significantly change the thermal conductivity of the lattice. This observation calls further studies of the fidelity of interatomic potentials where the cutoff distance is often chosen rather than determined from fundamental insights. Ab initio studies to remedy this issue should enable molecular dynamics to reliably predict properties, such as thermal conductivity, at the nanoscale and promote new material discovery.   

The discontinuous second-derivatives lead to jumps in elastic constants. This would prevent realistic studies of mechanical properties of semiconductors (in particular, the plastic deformation and the dislocation behavior). The proposed modification of the Tersoff potentials can therefore impact many problems including the mechanical properties of materials.

Sandia is a multi-program laboratory operated by Sandia Corporation, a Lockheed Martin Company, for the United States Department of Energy National Nuclear Security Administration under contract DEAC04-94AL85000. This work was performed under a Laboratory Directed Research and Development (LDRD) project. We are also grateful for helpful discussions with A. P. Thompson and G. J. Wagner.
 
\appendix
 

\end{document}